\documentclass[referee]{mn2e}
\usepackage{times,epsfig}
\newcommand{\aap}{A\&A}

\newcommand{\aj}{AJ}
\newcommand{\apj}{ApJ}
\newcommand{\apjl}{ApJL}
\newcommand{\apjs}{ApJS}

\newcommand{\mnras}{MNRAS}

\def \kms{\ifmmode{~{\rm km\,s}^{-1}}\else{~km~s$^{-1}$}\fi}
\def \vhel{\ifmmode{V_{{\rm hel}}}\else{$V_{{\rm hel}}$}\fi}
\def \vlsr{\ifmmode{V_{{\rm lsr}}}\else{$V_{{\rm lsr}}$}\fi}
\def \vsys{\ifmmode{V_{{\rm sys}}}\else{$V_{{\rm sys}}$}\fi}
\def \vobs{\ifmmode{V_{{\rm obs}}}\else{$V_{{\rm obs}}$}\fi}
\def \degree{\ifmmode{^{\circ}}\else{$^{\circ}$}\fi}
\def \lsun{\ifmmode{{\rm\ L}_\odot}\else{${\rm\ L}_\odot $}\fi}
\def \msun{\ifmmode{{\rm\ M}_\odot}\else{${\rm\ M}_\odot$}\fi}
\def \myr{\ifmmode{{\rm\ M}_\odot{\rm\ yr}^{-1}}\else{${\rm\ M}_\odot$ 
yr$^{-1}$}\fi}
\def \teff{\ifmmode{{\rm{T}}_{\rm eff}}\else{${\rm{T}}_{\rm eff}$}\fi}
\def \mdot{\ifmmode{{\rm\dot{M}}}\else{${\rm\dot{M}}$}\fi}

\newcommand{\ha}{H$\alpha$}

\newcommand{\oiii}{[O\,{\sc iii}]}
\newcommand{\oiiil}{[O\,{\sc iii}]\ 5007\,\AA}
\newcommand{\nii}{[N\,{\sc ii}]}
\newcommand{\niil}{[N\,{\sc ii}]\ 6584\,\AA}

\newcommand{\niiab}{[N\,{\sc ii}]\ 6548,\ 6584\,\AA}
\def \st{\ifmmode{^{\mathrm{st}}}\else{${^{\mathrm{st}}}$}\fi}
\def \nd{\ifmmode{^{\mathrm{nd}}}\else{${^{\mathrm{nd}}}$}\fi}
\def \rd{\ifmmode{^{\mathrm{rd}}}\else{${^{\mathrm{rd}}}$}\fi}
\def \th{\ifmmode{^{\mathrm{th}}}\else{${^{\mathrm{th}}}$}\fi}
\title[Helix planetary nebula]{Optical line profiles of the  
Helix planetary nebula (NGC~7293) to large radii.}
\author[J. Meaburn et al]
{J. Meaburn$^{1}$,  
J. A. L\'{o}pez $^{2}$ and
M. G. Richer$^{2}$.\\ 
$^{1}$Jodrell Bank Observatory, Dept of Physics \& Astronomy, University of
Manchester, Macclesfield, Cheshire SK11 9DL UK.\\
$^{2}$ 
Instituto de Astronom\'{\i}a, UNAM, Apdo. Postal 877,
Ensenada, B.C. 22800, M\'{e}xico.
}
\begin{document}

\date{Accepted yyyy mmmmmmmmmm dd. Received yyyy mmmmmmmmmm dd; in original 
form yyyy mmmmmmmmmm dd}

\pagerange{\pageref{firstpage}--\pageref{lastpage}} \pubyear{2004}

\maketitle

\label{firstpage}

\begin{abstract}
New, very long (25\arcmin), cuts of 
spatially resolved profiles of the \ha\ and \nii\
optical emission lines have been obtained over the face of the Helix
planetary nebula, NGC 7293. These directions were chosen to supplement
previous similar, though shorter, cuts as well as crossing
interesting phenomena in this nebular envelope. In particular, one new cut
crosses the extremes of the proposed CO J=2-1 emitting outer `torus'
shown by Huggins and his co--workers to be nearly orthogonal to its
inner counterpart. The second new cut crosses the  extensive outer filamentary
arcs
on either side of the bright nebular core.
It is shown that 
NGC~7293 is composed of multiple  bipolar outflows along different axes. 
Hubble-type
outflows over a dynamical timescale of 11,000 yr are shown to be occurring
for all of the phenomena from the smallest He{\sc ii} emitting core
out to the largest outer filamentary structure. All must then have been ejected
over a short timescale but with a range of ejection velocities.

\end{abstract}

\begin{keywords}
circumstellar matter: Helix Nebula: NGC 7293
\end{keywords}

\section{Introduction}

The Helix planetary nebula (NGC~7293) continues to attract both
observational and theoretical interest simply because it is one
of the closest (213 pc distant -- 
Harris et al. 1997) bright, evolved planetary nebulae
(PNe) and hence open to investigation on a wide range of spatial scales.
In both O'Dell, McCullough \& Meixner (2004) and Meaburn et al. (2005b),
 and references therein, 
much of the previous work is summarised up to those dates.
More recently, Hora et al. (2006) have made Spitzer Space Telescope 
infra-red observations where the knotty nature of the whole PN envelope
is spectacularly apparent in their images. 
Ultra-violet (UV) images of NGC~7293 released in 2005 and taken 
with the Galaxy Evolution Explorer
(GALEX) satellite reveal particularly well
many of the halo features including what was thought to be the large
filamentary bipolar lobe as well as a possible `jet' and the counter
bow-shaped feature previously noted in \niil\ emission by Meaburn et al.
(2005b). (The GALEX FUV (135-175 nm) image combined 
with the NUV (175-280 nm) one
can be examined in photojournal.jpl.nasa.gov/jpeg/PLA07902.jpg).
Theoretically, Dyson et al (2006)
have predicted the creation of the unambiguously accelerating tails
of the inner toroidal system of cometary knots as their dense core
are overrun by the mildly supersonic AGB `superwind'. Garcia-Segura et al
(2006) have also considered the effects on the PN envelope as the
highly supersonic `fast' wind switched off for this is no longer observed
in the spectrum of the central star (Cerruti--Sola \& Perinotto 1985).

The global structure of the complex PN envelope of NGC~7293 is becoming
clarified for the key to its understanding is knowledge of its 
large--scale kinematical behaviour. Initially, Meaburn \& White (1982)
from wide-field Fabry-Perot line profiles and imagery using the
\niil\ and \oiiil\ emission lines showed that the helical appearance
of the bright core of NGC~7293 is the manifestation of a bipolar
shell expanding at $\approx$ 25 \kms\ 
viewed at a 37\degree\ to its axis. 
This central bipolar structure
was also shown to contain an inner \oiiil\ emitting spherical shell
now (Meaburn et al 2005b) known to be expanding at only 12 \kms. Healey \&
Huggins (1990) and Young et al (1999) with complete coverage of
CO J=2--1 profiles over 
the bright, apparently helical structure refined this viewpoint and
furthermore suggested that the bipolar lobes  emanated from a 
central, clearly identifiable, torus expanding at 29~\kms.
Incidentally, this central structure, with 
two diametrically opposite, elongated,
expanding shells emanating from a central expanding torus will be
referred to throughout this paper as `bipolar' in keeping with
its previous description (see fig.11 in Meaburn et al. 2005). Overall it 
could equally be described as a quasi-ellipsoidal
expanding structure with a dense, toroidal waist. 
All features of this model
for the bright helical structure were consolidated by the 
morphological/kinematical modelling of the optical observations in
Meaburn et al. (1998 \& 2005b). The spatially resolved 
\niil\ profiles with the first Manchester
Echelle Spectrometer (MES -- Meaburn et al 1984) on the 
Anglo-Australian 3.9--m telescope
over 17\arcmin\ diameter east--west and north--south cuts (6 \& 7 in Fig. 1)
(Meaburn et al 1996 \& 1998) proved decisive in this analysis.

These previous long cuts of optical line profiles have now been supplemented
by even longer cuts at intermediate and strategically desirable position
angles over the nebular surface. Of particular interest are the anomalous 
velocity features in the CO J=2--1 profiles discovered by Healey \& Huggins
(1990) and mapped completely by Young et al (1999) which suggests that
there is a separate neutral torus, not seen hitherto in the imagery
or line profiles at optical wavelengths but which is both outside the
nebular core and expanding nearly orthogonally to it. One of the new extra-long
cuts of optical line profiles, to be reported here, crosses both the 
`point symmetric' CO features (see fig. 3 of Young et al 1999) that lead
to this orthogonal interpretation, as well as being orientated nearly down
the axis of the proposed `inner' bipolar lobe. The second new extra-long
cut crosses the main `helical' structure but extends continuously
out to the faint halo filaments on either side of this and
along the inner part of a possible jet. Only 
limited observations of the optical line profiles from 
the north-eastern halo filaments had been made previously (Walsh \& Meaburn
1987; Meaburn et al 2005b though note the calibration correction
made for the latter in the present paper).

The combination of these new observations with the previous long cuts
of optical line profiles now provides unprecedented kinematical
coverage of the principal features of NGC~7293. Consequently, 
the propensity in PNe
for the ejection of multiple bipolar lobes, each perpendicular to its own 
expanding torus themselves at various tilt angles,
is clearly revealed in NGC~7293.

 \section{Observations and Results}
\noindent 
The longslit observations were obtained with the Manchester Echelle
Spectrometer (MES--SPM Meaburn et al, 2003)
combined with the f/7.9 focus of the 2.1--m San Pedro M\'{a}rtir UNAM
telescope between 29 July and the 29 Aug. 2005.  
This echelle spectrometer has no
cross-dispersion. For the present observations, a filter of 90~\AA\
bandwidth was used to isolate the 87$^{th}$ order containing the \ha\ and
\nii\ nebular emission lines.

A SITE CCD with 1024~$\times$~1024 square pixels,
each with 24~$\mu$m sides, was
the detector.  Two times binning was employed in both the spatial and
spectral dimensions. Consequently 512 increments, each 0.624\arcsec\
long, gave a total projected slit length of 5.32\arcmin\ on the sky.
`Seeing' varied between 1-2\arcsec\ during these observations. 

Six overlapping slit positions were obtained along each of the
two cuts 4 and 5 in Fig. 1. These were each merged to form the
 single positional--
velocity (pv) arrays of \ha\ and  \niil\ profiles using the Starlink KAPPA 
CCDPACK MAKEMOS routines. Only the \niil\ arrays after this
process are shown in Figs 2 \& 3 for PA = 76\degree\ and 140\degree\ 
respectively for the \ha\
profiles are emitted by the whole of the internal expanding volumes of
NGC 7293 (Meaburn et al 2005b) and the final picture becomes confused.

The
slit was 150~$\mu$m wide ($\equiv$~11~\kms\ and 1.9\arcsec) for ten of the 
separate slit positions and 300~$\mu$m wide for only the two positions
at either end of cut 5.
Integration times were 1800 s in all cases and the spectra were calibrated in
wavelength to $\pm$~1~\kms\ accuracy when converted 
to heliocentric radial velocity (\vhel) against the spectrum 
of a Th/Ar arc lamp.

Incidentally, there is an error in the velocity scale of figs 3--5 in
Meaburn et al (2005b) for the pv arrays from
slit positions 1-3 (shown here in Fig. 1). The heliocentric correction
had not been applied to the velocity scale, consequently
the zero of the erroneous 
scales in these previous figures should be -26.1~\kms\ to
convert them to heliocentric radial velocity. This has been carried out
for the pv array of \niil\ profiles for slit position 2 in Fig. 1
and shown here in Fig. 5. The \niil\ profiles are compared in Fig. 5 with
the systemic heliocentric radial velocity \vsys\ = -27~$\pm$~2~\kms\
for the whole nebula (Meaburn et al 2005b - and see therein details
of these previous observations). Note that for comparison with the 
CO observations of Young et al (1999) their relationship 
\vhel\ = \vlsr\ - 3.2 \kms\ should be used though the STARLINK RV  routine
gives the difference as either -1.9 or -2.9 \kms.

\section{Discussion}

\subsection{The morphology of the large--scale features.}

Many of the large--scale optical 
phenomena are sketched in Fig. 6 as dark solid lines 
for the brighter features and dark dashed lines for the fainter ones. These
can be seen in the deep \ha\ + \niiab\ image 
presented in fig. 1a-c of Meaburn et al
(2005b) and spectacularly in the near UV (175--280 nm) Galaxy 
Evolution Explorer (GALEX) image (NASA/JPL--Caltech/SSC circulated in
2005). The faint nebulosity visible in the latter image 
may be a consequence of the dominant C{\sc iii}] 1906 \& 1909 \AA\ emission 
lines. The bright \niil\ emitting  central inner `ring' in Fig. 6 is on the
inside edge of a corresponding CO J=2-1 emitting ring (Huggins \& Healy 1986;
Healy \& Huggins 1990 and Young et al 1999) expanding at 29 \kms\
yet surrounds the system of slower moving cometary knots
whose expansion is only 14.5 \kms\ (Meaburn et al 
1996). 
Bipolar lobes project 
(Meaburn et al 1996\& 2005b) along the common axis of
these ionized/neutral tori.
The lobes of this central bipolar system  
manifest themselves as the bright helical filaments in Fig. 6. 

Young et al (1999) reveal clearly the presence of an outer CO expanding ring
whose curve of radial velocities (their fig. 5) shows that it is orientated
nearly orthogonally to its inner companion. The sides 
of this outer ring with its highest approaching and receding 
radial velocities are depicted
in Fig. 6 by the southeastern and northwestern 
hashed `CO' circles respectively. It is notable that 
the axis of this outer CO torus has approximately the same 
position angle (PA = 140\degree) as 
the faint extensions marked L1 and L2
in Fig. 6.

The jet--like feature and its possible bow--shaped filamentary counterpart
are also very clear in the UV GALEX image. The possible jet extends 
from inside the image of the filaments of the SW outer arc 
to the edge of the outer
envelope and because of the tentative identification of its nature
is marked jet? in Fig. 6. 

\subsection{The kinematics of the large--scale features}

 \subsubsection{Inner bipolar structure along PA = 125\degree.}

The detailed kinematical--morphological modelling in Meaburn et al (2005b) 
of the pv arrays
of \niil\ profiles along the previous cuts 6 \& 7 in Fig.1 
showed that the bright helical filaments and inner ring in Fig. 6
are the manifestations of an inner bipolar nebula with a central
expanding torus. 
In this model the axis of the central torus is taken to
be along PA $\approx$ 125\degree,
orientated at 37\degree\ to the sight line  and expanding at
14.5\kms\ i.e. the expansion velocity of the system of cometary knots. 
The lobes are expanding with Hubble--type velocities and reach
expansion velocities of 24 \kms. The southeastern and northwestern
lobes are tilted towards 
and away from the observer respectively. The tilt of the bipolar axis
and that of the central torus with respect to the sight line 
was determined simply 
by measuring the apparent dimensions of the ring (see Fig. 6) 
in optical images and by assuming the torus is circular.
 
This interpretation is consolidated in detail 
by the new profiles over
this core along cuts 4 \& 5 in Figs. 2 \& 3 respectively. The line 
profiles of the inner \niil\
emitting ring
of the central torus 
is revealed particularly well in Fig. 2, along cut 4, which
is aligned nearly with the bipolar axis at PA = 125\degree.
Here its emission manifests itself as the
two brightness maxima at 4\arcmin\ and -4\arcmin\ from the central
star at radial velocities of \vhel\ =
-36.7 $\pm$ 1 and -11.2 $\pm$ 1 \kms\ respectively.
These measurements were made by Gaussian fitting the single line
profiles extracted from those parts of the pv arrays containing
the maxima. 
The expansion velocity of the \niil\ emitting
torus can now be refined compared with the 14.5 \kms\ (the expansion
velocity of the system of cometary knots) used in the model in
Meaburn et al (2005b). With this direct measurement this becomes
12.75 (sin 37\degree)$^{-1}$ = 21.2 \kms. A systematic change in expansion
velocity throughout the inner neutral/ionized torus is then occurring with
the system of cometary knots nearest the central star expanding
at 14.5 \kms, the larger diameter intermediate 
\niil\ part of the torus (the `ring' in Fig. 6) at 21.2 \kms\ and 
the outermost CO component of this central torus (Young et al 1999) 
at 29 \kms.   

The change of expansion velocity of the central \niil\ emitting torus 
from 14.5 to 21.2 \kms\ in the
morphological/kinematical model in Meaburn et al (2005b) makes only
minor cosmetic differences to the predictions of the model for
comparison with the previous observations along cuts 6 \& 7 in Fig. 1.
In fact, the detailed comparison is improved quantitatively but will
not be repeated here.
The tilt to more
positive velocities from the southeastern to the northwestern
(bottom to top) is clear in Fig. 2 as predicted by the same bipolar/torus 
model. 

 \subsubsection{Outer CO torus}

Firstly, 
the outer CO ring appears
to have been detected at optical wavelengths for the first time
in the new pv array along cut 4. The very deep, negative, greyscale
image of this is presented in Fig. 4 and two sets of velocity `spikes'
(A and B in Fig. 4) can be seen over the approaching and receding maxima
depicted in Fig. 6. These have the same radial velocity ranges as the
corresponding CO features of Young et al (1999). The higher angular
resolution of the present \niil\ data compared with the CO maps (30\arcsec) 
appears
to have resolved this outer neutral/ionized torus into a double ring structure
with its axis 
along PA = 140\degree\ compared with PA = 125\degree\ for the inner torus 
(Sect.
3.2.1). There is no easy way to estimate directly the tilt of the axis of this
outer torus to the sight line for it does not appear on optical images
and most of its extent gets confused in the CO maps as it crosses the central
nebula, consequently an estimation of only a lower limit of its 
expansion velocity can be made i.e. $\geq$ 27 \kms. The possible
bipolar lobes L1 and L2 in Fig. 6 (Sect. 3.1.1) associated with this
outer torus would have receding and approaching radial velocities
respectively. Neither have yet been measured. Also the axis
of the outer CO torus and the common axis of these possibily
related bipolar lobes would expected to be the same.

 \subsubsection{Outer bipolar structure along PA = 50\degree.}

Evidence for some type of outer bipolar structure,
as suggested on morphological grounds in Sect. 3.1, enveloping the inner one, 
is also present in this new kinematical data. 
As the pv array crosses from the bright helical structure to the NE outer arc
in Fig. 3 along cut 5 the profiles stay on \vhel\ = -50 \kms
until they reach -95 \kms\ to then come back to \vsys\ = -27 \kms\
at the bright filamentary edge. Similar behaviour is shown in Fig. 5 
for the previous pv array along slit 2. Some sort of three dimensional
expansion at $\geq$ 68 \kms\ must be occurring with the bright NE 
outer arc filaments
being viewed tangentially through the edge of an expanding
shell. However, this partial (?) shell cannot have a simple, 
radially expanding, 
quasi-spherical structure for no receding velocities are detected
within its circumference.

The pv arrays in Fig. 3 for the southwestern end of cut 5 cover the 
filamentary feature that could be an inner extension of the `jet' in Fig. 6.
Within this complication, radial velocities are in receding directions 
as the SW outer arc is crossed which is consistent with the NE 
and SW outer arcs in Fig. 6 being part of the same, albeit not simple, bipolar
structure. 

\subsubsection{Jet and bow--shock}

Only very limited  information is in the present kinematical
data concerning the nature of the possible jet and bow--shock whose
morphology is discussed in Sect. 3.1. The most telling behaviour
is the Hubble--type increase in radial velocity along the possible
inner part of the jet as  shown along the southwestern end of the 
pv array in Fig. 3. The radial velocities in the knots in the `jet'
change systematically from \vhel\ = -10 \kms\ at about -8\arcmin\ from the
central star out to \vhel\ = 10 \kms\ at the bottom end of the array.
Unfortunately the jet-like feature nearest the outer envelope was not
covered in the present observations.

No direct kinematical information has yet been obtained over the bow--shaped
filament in the northwestern quadrant (Fig. 6). However, the
UV GALEX image reveals this to be the edge of extensive complex filamentary
structure up to the central bright helical filaments and the
outer arc may only be the edge of a three dimensional structure
typical of bow-shocks. In this case it could be possible
that the extreme velocity feature out to \vhel\ = -110 \kms\ 
at 6\arcmin\ from the central star along cut 5 in Fig. 3 is from this
more extended 
structure. More observations are needed to investigate this possibility
conclusively. 

\section{Episodic ejections within a  Hubble-type outflow.}

Further evidence is presented here that NGC 7293 is composed of at least two
bipolar nebulae, with their bipolar
axes  at substantially
different orientations along PAs = 125\degree\ 
and 50\degree\ respectively. The tilt of the axis of the 
the PA = 125\degree\  lobes are well- established as 37\degree\ with respect
to the sight line whereas that of the  PA = 50\degree\ lobes is as yet
unknown.
This multiple bipolar lobe structure is common in  other
`poly-polar' nebulae e.g. NGC 6302 (Meaburn \& Walsh 1980), KjPn~8
(L{\' o}pez, Vazquez \& Rodriguez 1995: L{\' o}pez et al 1997), 
NGC 2440
(L{\' o}pez et al 1998), J 320 (Harman et al 2004)
and see  Manchado et al
(1996).
 In fact this
multi-polar structure represents the rule in young proto--PNe (Su et al. 2003)
and the evolved PN NGC 7293 is most likely the natural
descendent of this early structure.

The radii and expansion velocities of the proposed bipolar outflows for
NGC 7293 are summarised
in Table 1 and plotted in Fig. 7. Also included are the radii and expansion
velocities
of the inner He{\sc ii} 4686\AA\ \& 6560\AA\ emitting volumes, the  inner
\oiii\ emitting shell and the system of cometary globules.
The most striking feature of the trends in Fig. 7 is that
Hubble--type expansion is occurring continuously from the
innermost volumes of NGC 7293 right out to the NE outer arc
(8 in Fig. 7). The simplest dynamical interpretation
of such behaviour is  that all features must have been formed
over a short period of time compared with the dynamical
age of the nebula,  in a sequence of ejection events, sometimes
along different axes; the fastest ejections in a particular direction
could  have simply travelled furthest and, if knotty (Hora et al. (2006),
separated out from those moving slower. This Hubble--type expansion
is not unexpected for it has been found in many younger PNe (Corradi 2004).
.

Most recently, from observations of the UV absorption line profiles
in the spectrum of the hot central star of the 
Dumbbell nebula (NGC 6853, M27) McCandliss et al (2007) and McCandliss \&
Kruk (2007) have consolidated in detail the same viewpoint 
(also see Meaburn et al 2005a) for this similarly
evolved PN. Wilson (1950) had come to the same conclusion for many
bright PNe. 

The present observations of the \nii\ radius and expansion velocity
(see 4 in Table 1)
of the inner torus of NGC 7293 
then offer a sound estimation of the dynamical age T${_D}$ = 
11,000 yr
of
the whole nebula for a distance of 213 pc (Sect. 1) and within the 
Hubble - type expansion assumption. This is consistent with the estimation
of T${_D}$ = 
10,000 yr for the inner CO emitting torus by Young et al (1999).

\begin{table*}
\centering
\caption{The expansion velocities for distinct components of NGC 7293 
are listed versus their radii.}
\begin{tabular}{|l|c|c|c|}
\hline
\multicolumn{4}{|c|}{}\\
\multicolumn{4}{|c|}{EXPANSIONS v. RADII} \\
\multicolumn{4}{|c|}{}\\
\hline
\hline
{\bf feature}&{\bf radius}&{\bf expansion velocity}&{\bf references}\\
1) He{\sc II} central volume &$\leq$ 1.6\arcmin\ &$\leq 12 \kms$ &O'Dell 
et al 2004: Meaburn 
et al 2005b\\
2) [O{\sc iii}] inner shell&2.2\arcmin& 12.5 \kms& Meaburn \& White 1982:
Meaburn et al 2005b \\
3) cometary globules&3.0\arcmin&14.5 \kms&Meaburn et al 1992\\
4) inner torus - [N{\sc ii}]&4.0\arcmin&21.2 \kms&present paper\\
5) inner torus - CO&4.3 \arcmin&29.0 \kms&Young et al 1999\\
6) inner torus -  [N{\sc ii}] lobes& 5 \arcmin& 24 \kms&Meaburn et al 2005b\\
7) outer torus  - CO& $\geq$ 3.8\arcmin& $\geq$ 27 \kms&Young et al 1999\\
8) NE outer arc&$\approx$ 12.5\arcmin&$\geq$ 68 \kms&present paper\\
\hline
\end{tabular}
\end{table*}

%\subsection{The interacting winds model}

%PNe are a consequence of the sequence of events as a star of $\leq$ 8 \msun\
%evolves through its Red Giant (RG) and  Asymptotic Giant Branch (AGB) 
%phases to finally become a White Dwarf (WD). In this sequence it is
%generally accepted that an initial  RG wind, is followed
%by the more volatile and many times denser 
%AGB `superwind' (with sporadic outbursts at 20 of \kms)
%which is then subjected to interaction with a 
%high--speed wind (several 1000 \kms)
%and becomes photo--ionised as the central star becomes a 
%WD with a high surface
%temperature. At this point the circumstellar envelope becomes the 
%embryo PN.
%The fast wind could then decline as the star becomes an older WD and the
%photoionised circumstellar 
%envelope would then be described as an evolved PN.

\begin{figure*}
\epsfclipon
\centering
\mbox{\epsfysize=4in\epsfbox[120 6 560 300]{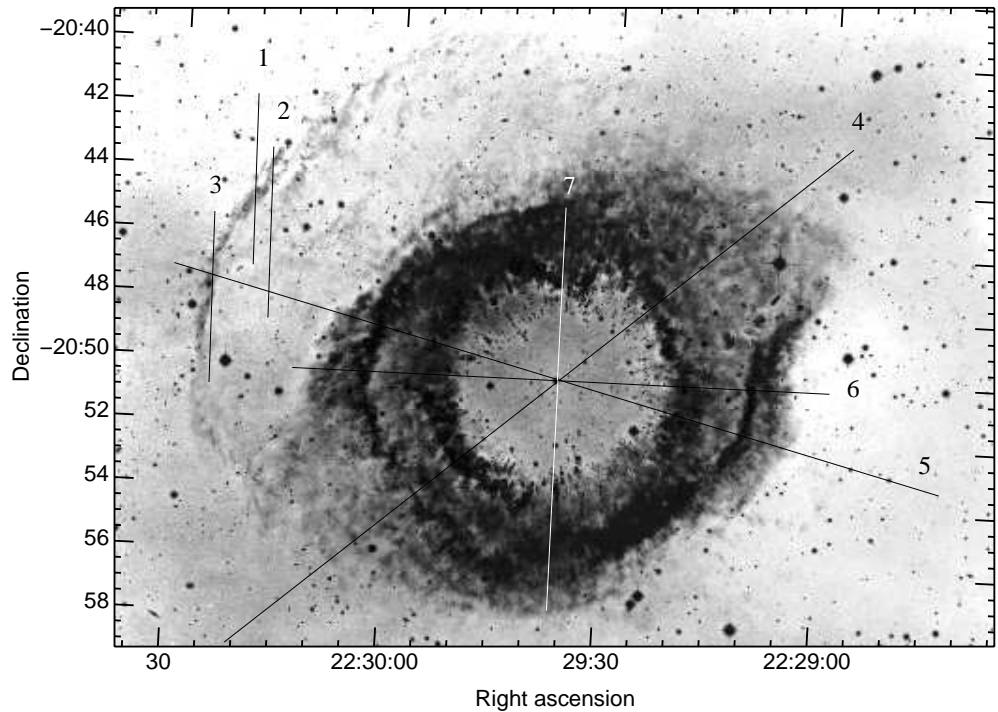}}
\caption{The previous slit positions and long cuts
of spatially resolved profiles 1, 2, 3, 6 and 7 and new ones,
4 and 5 are shown against the unsharp masked R image of NGC 7293
taken at the f/3.3 prime focus of the Anglo--Australian telescope by
Malin (1982). The coordinates are 2000 epoch.}
\label{reffig1}
\end{figure*}

\begin{figure*}
\epsfclipon
\centering
\mbox{\epsfxsize=4in\epsfbox[0 0 540 754]{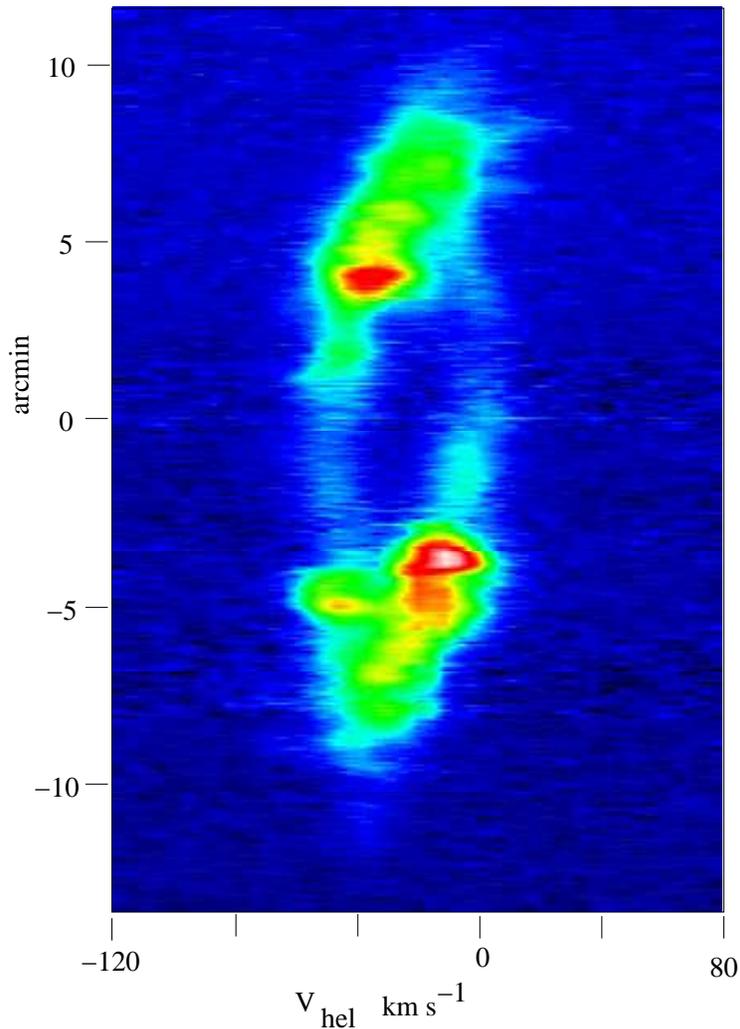}}
\caption{A positional--velocity display of \niil\ line profiles along the new
cut number 4 in Fig. 1 is shown. The vertical axis is the offset from the
central star (the horizontal continuous spectrum) and the south eastern
quadrant is to the bottom. This array is composed of the merged spectra
from six separate but overlapping slit positions.
The surface brightnesses have been converted to their  natural logarithmic 
values
and the colours used in the display are equally separated on this natural
logarithmic 
scale.}
\label{reffig2}
\end{figure*}

\begin{figure*}
\epsfclipon
\centering
\mbox{\epsfysize=6in\epsfbox[0 0 540 754]{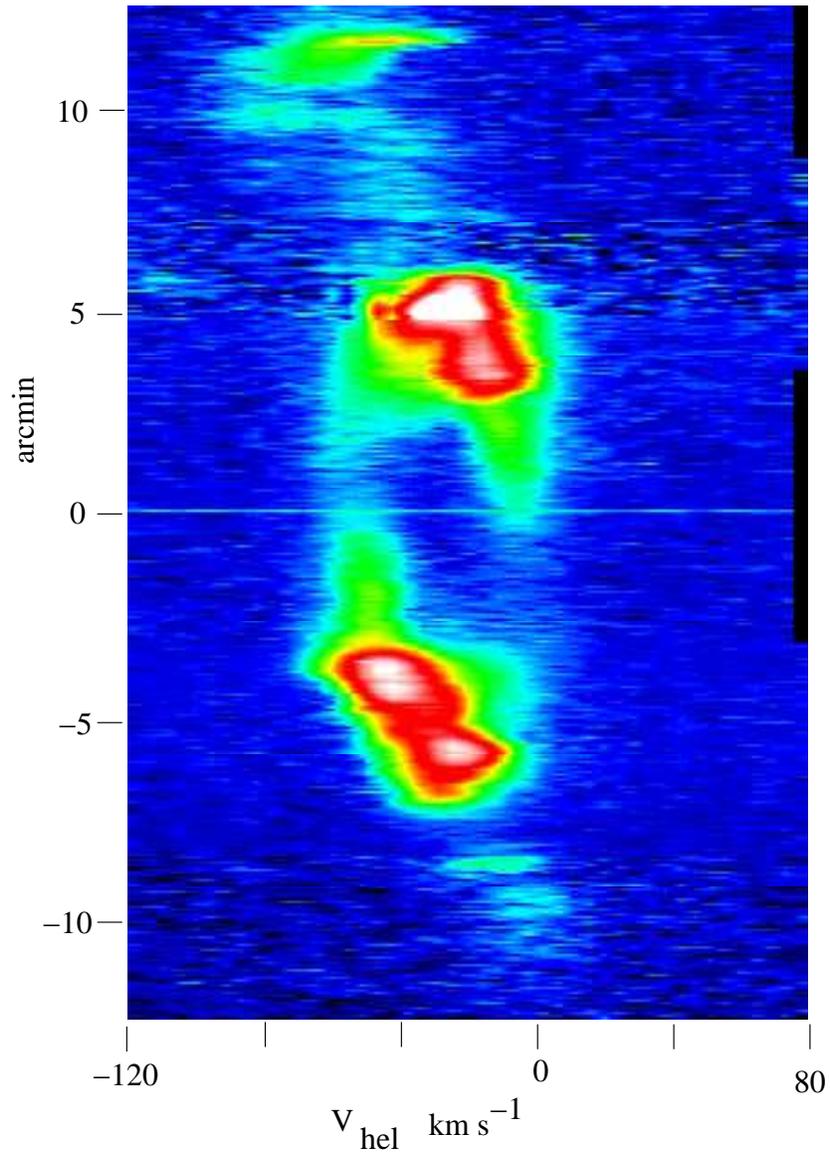}}
\caption{As for Fig. 2 but for cut 5 in Fig. 1. The south--western
quadrant is to the bottom.}
\label{reffig3}
\end{figure*}

\begin{figure*}
\epsfclipon
\centering
\mbox{\epsfysize=4in\epsfbox[0 0 523 766]{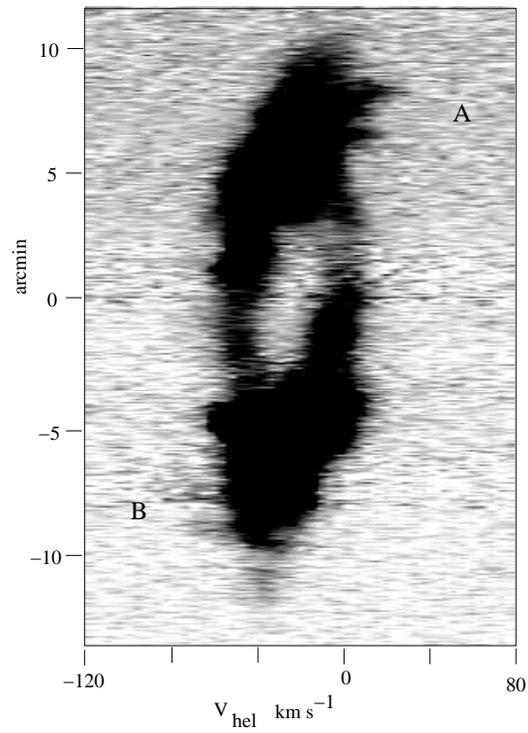}}
\caption{A very deep, negative, greyscale representation of the same
positional velocity array as shown in Fig. 2. The velocity `spikes'
marked A and B match in position and extent the comparable CO features.
The faint \niil\ 
profiles across the south--eastern halo can also be seen at the
bottom of the display.}
\label{reffig4}
\end{figure*}

\begin{figure*}
\epsfclipon
\centering
\mbox{\epsfysize=4in\epsfbox[0 0 523 766]{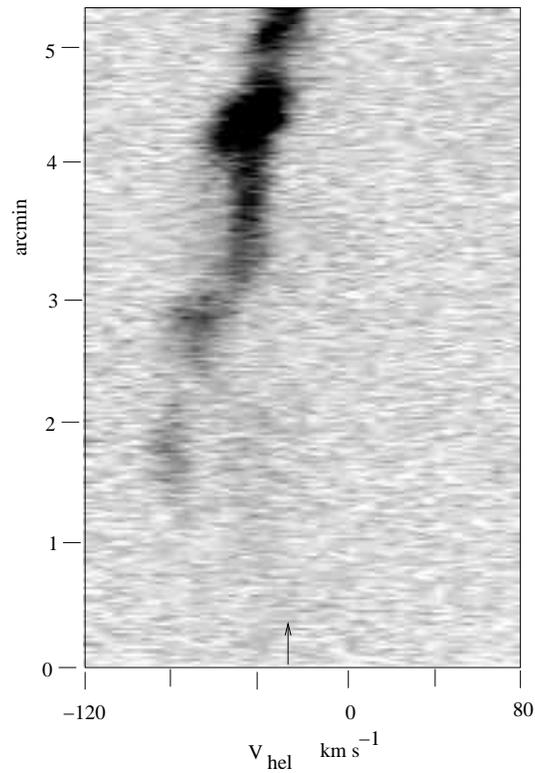}}
\caption{The positional-velocity array of  \niil\ profiles from the 
previous slit position 2 in Fig. 1 is presented now with the
corrected \vhel\ (see text). This array 
is compared with the systemic heliocentric
radial velocity (arrowed) of the whole nebula.}
\label{reffig5}
\end{figure*}

\begin{figure*}
\epsfclipon
\centering
\mbox{\epsfysize=5in\epsfbox[83 0 420 257]{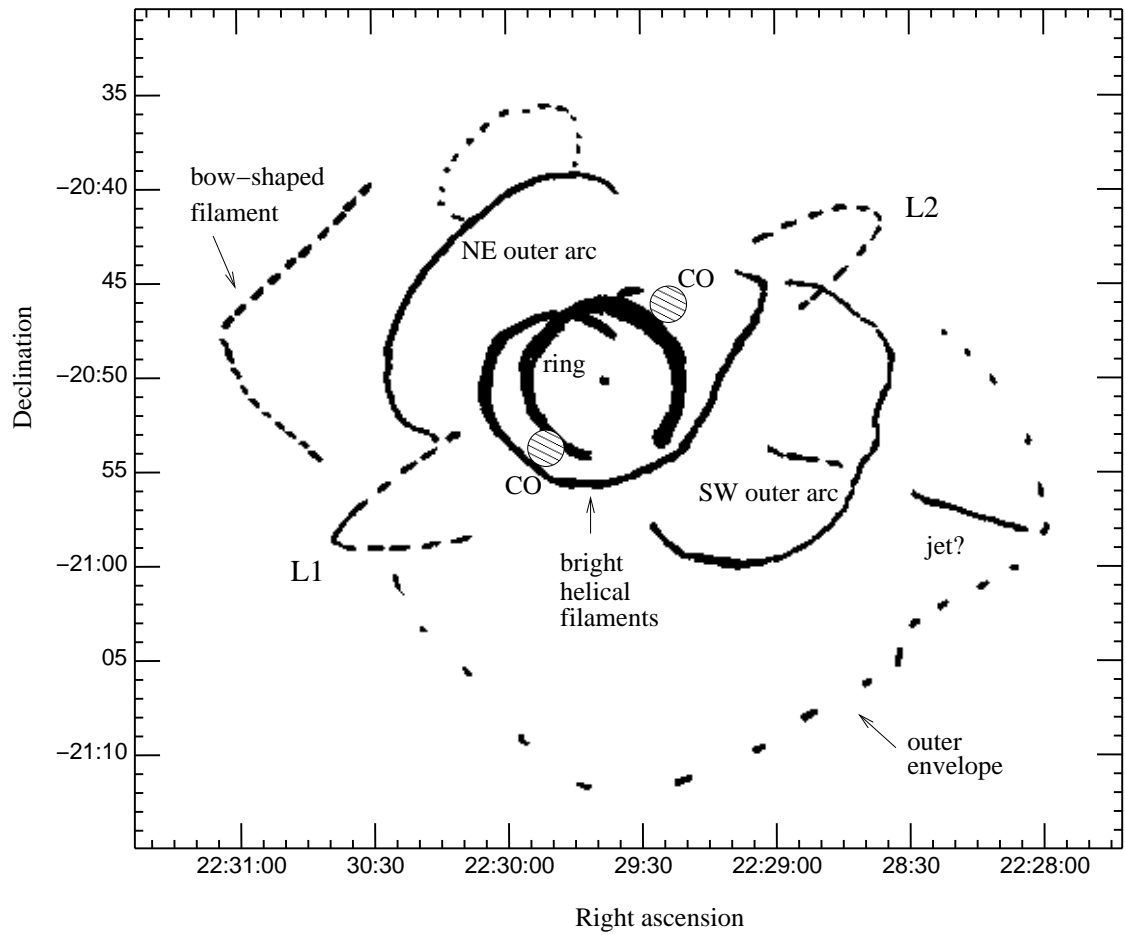}}
\caption{The salient optical and ultra--violet 
emission features of the large-scale
structure of NGC~7293 are compared with the approaching (-50 \kms\ -- hashed
circle to the SE) and receding (+4 \kms\ -- hashed circle to the 
NW) parts of the
outer CO torus discovered by Young et al (1999). The complete outer 
torus is not detected
in the CO or UV images as it gets confused with the brighter 
nebular features. The possible bipolar lobes emanating from 
this outer torus are 
L1 and L2 in which case they would have receding and approaching radial
velocities respectively and would have to have a common axis with it.}
\label{reffig6}
\end{figure*}

\begin{figure*}
\epsfclipon
\centering
\mbox{\epsfysize=6in\epsfbox[0 0 541 503]{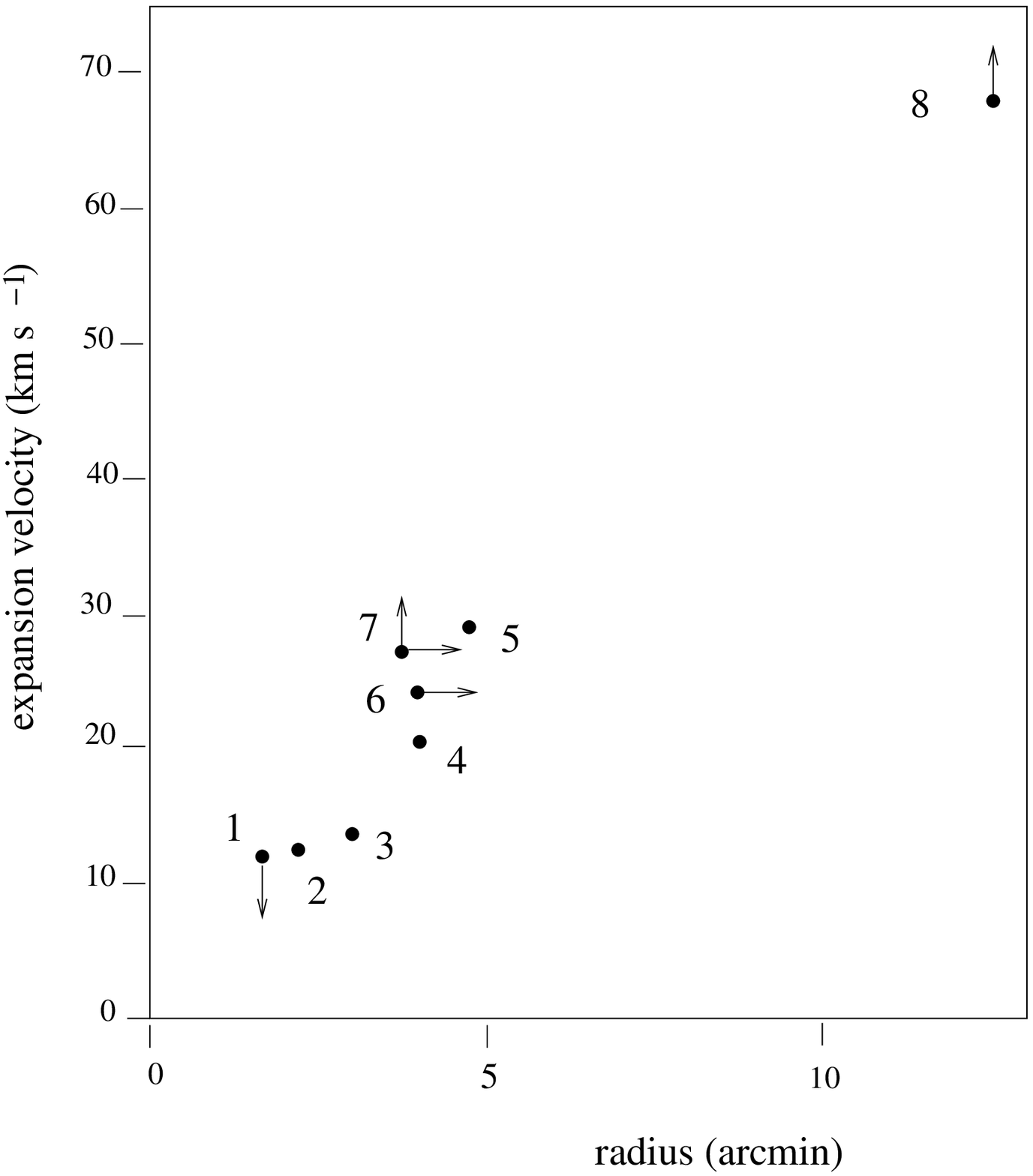}}
\caption{The radial expansions versus radii for the features of
NGC 7293 listed as 1 to 8 in Table 1 are shown. Arrowed lines indicate where 
values are either upper or lower limits for both axes. In the case
of 1 i.e. He{\sc ii}6560\AA (Meaburn et al 2005b) the profile is distinctly
top-hatted but the separate velocity 
components not quite resolved. For 7 the tilt
to the sight line is unknown affecting both the estimation of the radius
and the expansion velocity whereas for 8 the radius is reasonably
obtained from the optical image in Fig. 1 but the tilt of the
proposed bipolar axis is unknown hence the lower limit 
for the expansion velocity.}
\label{reffig7}
\end{figure*}

\section*{Acknowledgements}
JM is grateful to UNAM for supporting his 2007 visit to the Instituto
de Astronom\'{i}a, Ensenada where this paper was initiated
and to Myfanwy Lloyd (n{\'e} Bryce) for advice on the use of the MAKEMOS
Starlink routines. JAL and MGR
are in grateful receipt of DGAPA - UNAM grants IN 112103, 108406-2 and 108506.

\label{lastpage}
\end{document}